\documentclass{emulateapj}
\usepackage{graphicx}
\def\hmpc		{{~$h^{-1}$~Mpc}}
\def\ctunit             {{$^\circ$~$l^{-1}$}}

\begin{document}
\title{Crawling the Cosmic Network: Identifying and Quantifying Filamentary Structure}
\author{Nicholas A. Bond\altaffilmark{1}, Michael A. Strauss, Renyue Cen}
\affil{Princeton University}
\affil{Princeton University Observatory, Princeton, NJ 08544\label{Princeton}}
\altaffiltext{1}{nbond@physics.rutgers.edu}

\begin{abstract} 

  We present the Smoothed Hessian Major Axis Filament Finder (SHMAFF), an algorithm that uses the eigenvectors of the Hessian matrix of the smoothed galaxy distribution to identify individual filamentary structures.  Filaments are traced along the Hessian eigenvector corresponding to the largest eigenvalue, and are stopped when the axis orientation changes more rapidly than a preset threshold.  In both N-body simulations and the Sloan Digital Sky Survey (SDSS) main galaxy redshift survey data, the resulting filament length distributions are approximately exponential.  In the SDSS galaxy distribution, using smoothing lengths of $10$\hmpc\ and $15$\hmpc, we find filament lengths per unit volume of $1.9 \times 10^{-3}$~$h^2$~Mpc$^{-2}$ and $7.6 \times 10^{-4}$~$h^2$~Mpc$^{-2}$, respectively.  The filament width distributions, which are much more sensitive to non-linear growth, are also consistent between the real and mock galaxy distributions using a standard cosmology.  In SDSS, we find mean filament widths of $5.5$\hmpc\ and $8.4$\hmpc\ on $10$\hmpc\ and $15$\hmpc\ smoothing scales, with standard deviations of $1.1$\hmpc\ and $1.4$\hmpc, respectively.  Finally, the spatial distribution of filamentary structure in simulations is very similar between $z=3$ and $z=0$ on smoothing scales as large as $15$\hmpc, suggesting that the outline of filamentary structure is already in place at high redshift.

\end{abstract}


\clearpage
\section{Introduction}

Observational evidence for filamentary structures in the large-scale distribution of galaxies was first presented in galaxy redshifts surveys \citep[e.g.][]{TG78,Davis82,CfaSurvey,Sathy98, 2dF, SloanGreatWall}.  When similar structures were seen in cosmological N-body simulations of the dark matter distribution \citep[e.g.][]{CosmicWeb, Sathy96, MMF, Hahn07a}, a picture of a vast `cosmic web,' in which filaments skirted the boundaries of voids and were connected by galaxy clusters, began to emerge.  These filaments are thought to provide pathways for matter to accrete onto galaxy clusters \citep[e.g.][]{CL0016} and to torque dark matter halos to align their spin axes \citep{Hahn07a,Hahn07b,Hahn09}.  Filaments also produce deep potential wells and will give rise to a gravitational lensing signal on the largest scales \citep{LensingFils2, LensingFils}.  A number of authors have claimed detections of filaments using weak lensing \citep[e.g.][]{Kaiser98,LensingFils2,LensingFils}, but simulations predict that structure along the line of sight should produce shear comparable to that of the target filaments \citep{Dolag06} and the evidence remains far from conclusive.  In addition, the formation of filaments is accompanied by gravitational heating, which gradualy raises the temperature of the intergalactic medium over time and produces the so-called warm-hot intergalactic medium (WHIM) by $z=0$ \citep[e.g.][]{CO99}.

Perhaps the simplest and most effective means of identifying {\it clusters} in discretely sampled fields, such as redshift surveys and N-body simulations, is the friends-of-friends algorithm \citep[FOF,][]{FOF}, in which particle groups are assembled based on the separation of nearest neighbors.  These FOF structures can then be quantified with `Shapefinders,' statistics which measure the length, breadth, and thickness of structures and are related to the Minkowski functionals \citep{Shapefinders}.  \citet{SURFGEN} have developed an algorithm for computing the Shapefinders on structures at an arbitrary density threshold.  Many of those found in data and simulations are indeed filamentary, but FOF algorithms are optimized for structures that lie above a set density threshold, a condition approximately met by clusters at the present epoch.  Filaments and walls, however, are not bound and a strict density cut alone would not provide clean samples of such structures.

Another algorithm, called the Skeleton \citep{Skeleton,SkeletonData,Skeleton3D}, identifies filaments by searching for saddle points in a density field and then following the density gradient along the filament until it reaches a local maximum.  Although it appears to be effective at making an outline of the cosmic network, it lacks an intuitive definition of filament ends.  \citet{SpineWeb} also lacks such definition, but has been successful at tracing the filament network in cosmological simulations using watershed segmentation \citep[see also][]{Platen07} and a Delaunay tessellation density estimator \citep{Schaap00}.  If we wish to analyze filament length distributions or their spatial relationship to clusters,  it is important to separate individual filaments in the cosmic web.  Structure-finding techniques that only detect filaments between galaxy cluster pairs \citep[e.g.][]{Pimbblet05a,Colberg05,GP09} would present a biased view of the filament-cluster relationship.

An early technique for identifying filaments in two-dimensional data was developed by \citet{GottFil} that works on a similar principle to the algorithm described in this paper.  It divides the density field into a pixelized grid and identifies as filament elements any grid cell that has a larger density than its immediate neighbors along two of the four axes (including the two coordinate axes and two axes at $45^\circ$~angles to the grid) through the grid cell.  The algorithm was run on the Shane-Wirtanen galaxy count catalogue \citep{ShaneWirtanen}, but has not been developed further.  A later algorithm, presented by \citet{Dave97}, works on a similar principle, identifying ``linked sequences'' using the eigenvectors of the inertia tensor.  The authors found that the algorithm was poor at discriminating between cosmological models using CfA1-like mock galaxy catalogs, but primarily because of the small number of galaxies in the catalogs.

In Paper~$1$, we used the distribution of the Hessian eigenvalues of the smoothed density field ($\lambda$-space) on a grid to study three types of structure: clumps, filaments, and walls.  Filaments were found in the $\lambda$-space distributions at a variety of smoothing scales, ranging at least from $5-15$\hmpc, in both N-body simulations and the galaxy distribution measured by the Sloan Digital Sky Survey \citep[SDSS,][]{SDSS}.  Furthermore, filaments were found to dominate the large-scale distribution of matter using smoothing scales of $10-15$\hmpc, giving way to clumps with $\sim 5$\hmpc\ smoothing.

The fact that the eigenvalues of the Hessian can be used to discriminate different types of structure in a particle distribution is fundamental to a number of structure-finding algorithms \citep[e.g.,][]{CPS00,Hahn07a,MMF,FR08}. However, the relationship between $\lambda$-space and a particular structure is not always trivial.  For example, one might think that a filamentary grid cell would have two positive and one negative eigenvalue.  This will be true near the centre of a filament connecting two overdense filament ends, but in the vicinity of the overdensities or in the case that the filament ends at an underdensity, all three eigenvalues will become negative.  In addition, when working with a smoothed density field, these criteria select regions that are near clumps and do not necessarily lie along the filament.  Finally, these criteria disregard the structure's width -- for example, the regions away from the centre of the filament may have positive values of $\lambda_2$.

In this paper, we will describe a procedure to identify filaments in the three-dimensional galaxy distribution using an algorithm called the Smoothed Hessian Major Axis Filament Finder ({\small SHMAFF}), and compare their properties in cosmological N-body simulations to those in the SDSS galaxy redshift survey.  We describe our methodology, which uses the eigenvalues and eigenvectors of the smoothed Hessian matrix \citep[see ][hereafter, Paper~$1$]{paper1}, in \S~\ref{subsec:Method}.  In \S~\ref{subsec:FilPars}, we run the code with a range of possible input parameters and justify our choices for each.  We discuss the behavior of the algorithm when used on Gaussian random fields in \S~\ref{subsec:FilGauss}, allowing us to distinguish those features of the large-scale distribution of matter that are a direct consequence of the non-linear growth of structure.  In \S~\ref{subsec:MockFils}, we use mock galaxy catalogues to estimate the incompleteness and contamination rates of filament samples and then use these quantities to interpret the distribution of filaments found in the SDSS (\S~\ref{subsec:FilData}).  In \S~\ref{subsec:Results} we summarize our results and discuss the implications of our findings.

\section{  Finding individual filaments   }
\label{subsec:Method}

Filaments, clusters, and walls all present sharp features in the density field along at least one of their principal axes.  In Paper 1, we described a procedure to generate a matrix of Gaussian-smoothed second derivatives of the density field (the Hessian matrix) at each grid cell, computing its eigenvalues, $\lambda_i$ (defined such that $\lambda_1<\lambda_2<\lambda_3$), and eigenvectors, $A_i$.  For the testing and development of the algorithm, we ran a series of cosmological $N$-body simulations, using a particle-mesh code with $\Omega_m = 0.29$, $\Omega_\Lambda = 0.71$, $\sigma_8 = 0.85$, and $h = H_0/(100 $~km~s$^{-1}$~Mpc$^{-1}$)$=0.69$ (see Paper 1 for details).  The simulation is performed within a $200$\hmpc~box with $512^3$ particles, each with mass, $m_p = 4.77 \times 10^9$~$h^{-1}$~$M_{\sun}$.

\begin{figure}[t]
\plotone{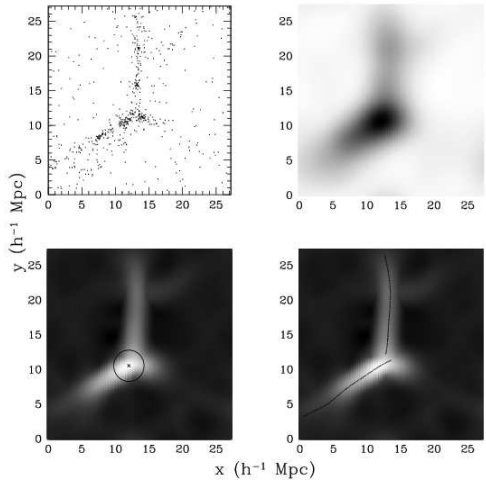}
\caption{ Slice from a cosmological simulation $10$\hmpc~deep and
$27.21$\hmpc~on a side, encompassing a prominent filamentary
structure.  Shown are the galaxies (upper left), density map (upper
right), and $\lambda_1$ map (lower two panels), where smoothing is performed on a scale of $l=2$\hmpc.  The cross in the lower left panel indicates the minimum value of $\lambda_1$ on the slice.  This will be the
starting point for the first filament traced by the algorithm.  The
circle around the cross has a radius equal to the removal
width (see text).   The solid lines in the lower right panel indicate the filaments as traced by a 2D version of SHMAFF.\label{fig:OneFil}}
\end{figure}

In order to generate a three dimensional distribution of mock galaxies, we first identify dark matter halos within the particle distribution using the HOP algorithm \citep{HOP} and then populate them using the halo occupation distribution and parametrization of \citet[][see Paper~$1$ for details]{HODPars2}. The resulting mock galaxy distribution is smoothed using a Gaussian kernel and its second derivatives, yielding a $128 \times 128 \times 128$ grid with Hessian eigenvalues and eigenvectors in each cell.  In Fig.~\ref{fig:OneFil}, we plot a slice from the simulation $10$\hmpc~deep and $27.21$\hmpc~on a side, chosen to encompass a prominent filamentary structure.  Shown are the galaxies (upper left), galaxy density map (upper right), and $\lambda_1$ map (lower left and right), smoothed with a $l=2$\hmpc~kernel to bring out the filament.  The structure appears most clearly in $\lambda_1$, so we construct a list of grid cells, G, ordered by increasing value of $\lambda_1$.  Before marking the first filament, we remove from G all grid cells that satisfy any of the following criteria, \begin{eqnarray} \label{eq:CellRemove}
 \lambda_1>0  \nonumber \\
 \lambda_2>0  \\
\rho<\bar{\rho}, \nonumber
\end{eqnarray}  
where $\bar{\rho}$ is the mean density of objects making
up the density field.  The $\lambda_1$ and $\lambda_2$ thresholds follow
from the definition of a filament -- the density field must be 
concave down along at least two of the principal axes.

The first element in G (the most negative in $\lambda_1$) is marked with a cross in the lower left panel of Fig.~\ref{fig:OneFil}.  From this starting point, we trace out the filament in both directions of the `axis of structure' (parallel and antiparallel to $A_3$), taking steps equal to the grid scale of $1.5625$\hmpc.  Subsequent filament elements are not constrained to lie on the grid, so we use a third-order polynomial interpolation scheme {\citep{NumericalRecipe}} on the grid to obtain the local Hessian parameters.  If, at any point along the filament, the angular rate of change of the axis of structure exceeds a threshold, $C$, we stop tracing and mark the point as a filament end.  The stopping condition at step m is given by,
\begin{equation}
\label{eq:AngleCrit}
\left|\mathbf{A_3}_{,{\mathrm m}} \times \mathbf{A_3}_{,{\mathrm m}-1}\right|>\sin(C\,\Delta),
\end{equation}
where $\Delta$, the grid cell size, is also the size of each step.  The filament finder will also stop and mark a filament end if it passes into a cell that satisfies one or more of the criteria specified in Equation~\ref{eq:CellRemove}.  In the lower right panel of Fig.~\ref{fig:OneFil}, we show the filaments that result from a sample run of SHMAFF on a $27 \times 27 \times 10$~$h^3$~Mpc$^{-3}$ slice from the dark matter particle distribution in our cosmological simulation.  

For each step along a filament, all grid cells within a removal width, W,
of the most recently chosen filament element, are removed from G, where 
\begin{equation}
\label{eq:RemoveWidth}
W_{\mathrm i}=K\sqrt{\frac{-\rho_{\mathrm i}}{\lambda_{\mathrm{1,\,i}}}}.
\end{equation}
In order to avoid tracing a filament more than once, subsequent
filaments cannot start within one of the removed cells.  They may,
however, extend into a removed cell, so long as the cell is not
excluded by any of the criteria given in Equations~\ref{eq:CellRemove}
and \ref{eq:AngleCrit}.  For a cylindrical filament with a Gaussian
cross section extending into a zero-density background, a value of
$K=1$ should exclude those parts of the structure that are not already
excluded by Equation~\ref{eq:CellRemove}.

The filaments traced by the above algorithm may be
offset from the ridges in the initial point field because of the finite
resolution of the grid.  Thus, we adjust the position of a
filament element, j, based on the average {\it perpendicular displacement}
of nearby grid cells from the filament axis, 
\begin{equation} 
\label{eq:FilCenter}
{\mathbf{\bar \Delta
s}}_{\mathrm j}=\frac{\sum_{\mathrm i=1}^{N}\mathbf{R}_{\mathrm i}}{N}, \end{equation}
where, 
\begin{equation} \label{eq:PerpDisplace}
\mathbf{R}_{\mathrm i}=\mathbf{\hat{A}}_{\mathrm j}\times(\mathbf{\hat{A}}_{\mathrm j}\times(\mathbf{x}_{\mathrm j}-\mathbf{x}_{\mathrm i})).
\end{equation} 
Here, $\mathbf{\hat{A}}_{\mathrm j}$ is the unit vector along the axis of structure (with arbitrary sign) and N is the number of objects in the initial point field that are within a smoothing length.  Application of the centering algorithm can result in fragmented filaments when shot noise is non-negligible, so we will not run it on point distributions with very sparse sampling, such as the SDSS galaxy distribution.

\begin{figure}[t]
\plotone{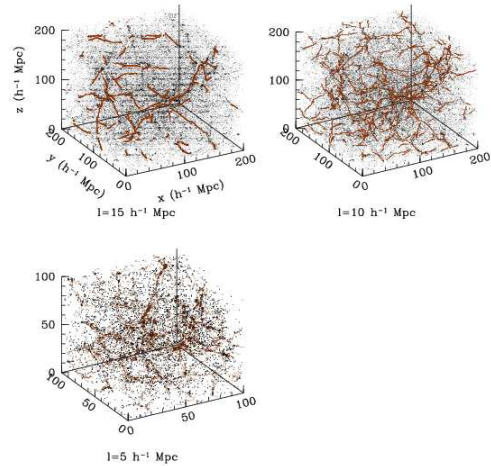}
\caption{Filaments found in the $z=0$ dark matter distribution of a 
cosmological simulation are plotted (in red) over a subsample of dark
matter particles (black).  Filaments are found in a density field
smoothed with the kernel length indicated below each
box.  For $l=5$\hmpc, we only show an octant of the full simulation box, as the full filament distribution is so rich that the figure for the full box would be too crowded.\label{fig:FilSmlens}}
\end{figure}

\section{ Filament-finding parameters   }
\label{subsec:FilPars}

In the filament-finding routine described in \S~\ref{subsec:Method}, there are two free parameters, the curvature criterion for identifying the filament ends, $C$, and the width of filament removal, $K$ (see Equation~\ref{eq:RemoveWidth}).  In principle, the optimal values of these parameters can be functions of the smoothing scale, redshift, or type of tracer (e.g., galaxies, dark matter particles), so it is important to understand their impact on the algorithm's output.  In this section, we will test the performance of the code on the distribution of dark matter particles in our cosmological simulation as a function of $K$, $C$, and the sampling rate.

In Fig.~\ref{fig:FilSmlens}, we show a run of the filament finder on the $z=0$ dark matter distribution of the cosmological simulation, using $C=40$\ctunit~and $K=1$.  Output is shown for smoothing with $l=15$, $10$, and $5$\hmpc, illustrating the scale-dependence and coherence of the cosmic network.  Any given filament will be found on a range of scales, depending on its width and length, but as the smoothing length is made smaller, the filament will be broken up into substructures which will themselves be filamentary or clump-like (see figures~15 and 16 in Paper~1).

\begin{figure}[t]
\plotone{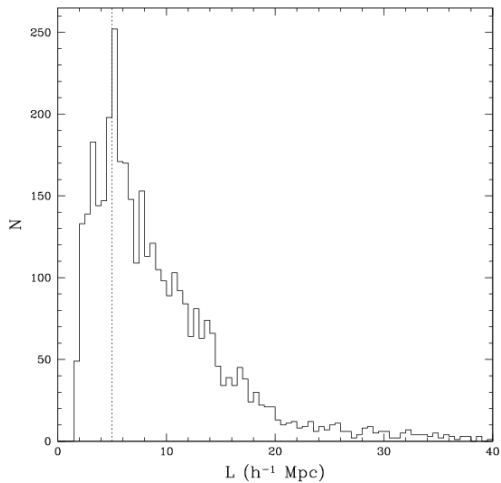}
\caption{Length distribution for a sample of filaments found
in the $z=0$ smoothed dark matter distribution (smoothing kernel width
of $l=5$\hmpc).  The dashed line indicates the smoothing length, below which filaments are removed from the sample.\label{fig:LengthDistroCut}}
\end{figure}

\subsection{ Sampling the filament length distributions }
\label{subsubsec:LengthDistroCut}

Before defining the parameters that are used to find filaments, we must decide on what we are willing to accept as a real filament.  An isolated spherical overdensity should not be viewed as a filament, but the filament finder would treat it as a very short ridge, tracing it from its centre until random fluctuations caused the axis of structure to deviate more than $C$, producing a `short filament'.  Fig.~\ref{fig:LengthDistroCut} shows the raw distribution of filament lengths for our dark matter simulation shown in Fig.~\ref{fig:FilSmlens}, using $C=30$\ctunit, $K=1$, and a smoothing length of $5$~\hmpc.  Not surprisingly, the distribution exhibits a dramatic drop-off below a smoothing length.  With this in mind, we hereafter discard filaments whose lengths are shorter than the smoothing length as non-physical.

\subsection{The $C$ parameter}
\label{subsubsec:Ct}

The traditional picture of large-scale structure as a `cosmic
web' \citep{CosmicWeb} suggests that filaments are
connected, one-dimensional strands that end abruptly at their points
of intersection.  As one filament begins and another ends, the local
axis of structure should change direction rapidly.  The $C$ parameter
denotes the maximum angular rate of change in the axis of structure
along a filament.  If this threshold is exceeded, filament tracing is
stopped.

In order to test the sensitivity of the output filaments to the value of the $C$ parameter, we set $K=1$ and generated filament networks in the N-body simulation with a range of $C$.  In all of these tests, increasing the value of $C$ led to an increase in the average length of the filaments and a decrease in the total number of filaments found.  If the curvature criterion is not strict enough, a filament will be traced past its vertex and into another filament.  Since our algorithm only prevents filaments from starting within previously-identified filaments (they are allowed to cross one another), this can lead to double detections of filaments.  We can obtain a rough count of these double detections by comparing filament elements to one another, where a filament element is defined as a single step (of interval, $\Delta$) on the grid.  In other words, for each step along a given filament, we find the closest filament element that is not a member of that same filament.  If the closest filament element is within a smoothing length and has an axis of structure within $C$, then the original element is labelled a `repeat detection.'  The total number of repeat detections in an output filament network is denoted by $R$.  The total length of the network at this scale is therefore given by
\begin{equation}
\label{eq:LengthTotal}
L_f=(N_e-R)\Delta,
\end{equation}
where $N_e$ is the total number of filament elements found and $\Delta$ is the step size taken by the filament finder.  Non-filamentary regions of space have already been excluded by the criteria in Equation~\ref{eq:CellRemove}, so an optimum set of parameters will maximize $L_f$ while minimizing $R$.

\begin{figure}[t] 
\plotone{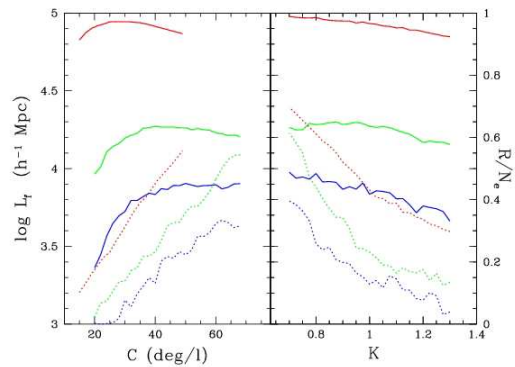}
\caption{Total length of the output filament network (solid lines) and the fraction of `repeat detections' (dashed lines) as a function of the curvature criterion, $C$ (left) and $K$ (right), The value of $C$ determines the filament ending points and the value of $K$ determines the distance from filaments that pixels are removed from further consideration by the filament finder.  Filaments were identified on three different smoothing scales, $l=15$\hmpc~(blue), $l=10$\hmpc~(green), and $l=5$\hmpc~(red).  The total length of the network (after removing repeat detections) maximizes at a value of $C$ that depends on smoothing length.\label{fig:Crit1Tots}}
\end{figure}

In the left panel of Fig.~\ref{fig:Crit1Tots}, we plot both the
fraction of repeat detections ($R/N_e$, dashed lines) and the total
length of the network ($L_f$, solid lines) as a function of $C$.  On
all smoothing scales, the fraction of false positives rises steadily
with increasing $C$, with no obvious breaks or minima.  The total
length, however, tends to rise until it reaches a maximum, after which
point it either flattens or falls slowly.  This suggests that, as long
as the curvature criterion is above a critical value, the algorithm will
trace out the entire filament network.  Since the fraction of false
positives rises with $C$, we will hereafter use a curvature criterion
near this value; that is, $C=50$, $40$, and $30$\ctunit~for
$l=15$, $10$, and $5$\hmpc, respectively.

\subsection{The $K$ parameter}
\label{subsubsec:KW}

As each filament is found, we wish to remove from the grid as much of it as possible without preventing the detection of further real filaments.  Using the previously-determined critical values of $C$, we ran the filament-finder with a range of $K$ and computed the total length of the filament network and the fraction of repeat detections as a function of $K$.  The results are shown in the right panel of Fig.~\ref{fig:Crit1Tots}.  All of the curves are monotonic, with repeat detections and the network length decreasing with increasing $K$.  Hereafter, we will set $K=1$ because it yields $R/N_e\lesssim 20$~per~cent.

\subsection{ Effects of sparse sampling}
\label{subsec:SparseFils}

In real galaxy catalogues, the number of galaxies per smoothing volume will sometimes be small and it is important to understand the impact of shot noise on the algorithm's ability to trace the filament network.  In a density field with sparse sampling, shot noise will create spurious filament detections in addition to the `repeat detections' described in \S~\ref{subsubsec:Ct}.  We have an effectively shot-noise-free density field in the dark matter particle distribution (with the simulation using a mean dark matter particle density of $17$~particles~$h^3$~Mpc$^{-3}$), so we perform sparse sampling on this field and use the complete particle distribution as a standard for comparison.  We construct three such data sets, sampled to densities of $5 \times 10^{-3}$, $2 \times 10^{-3}$ and $1 \times 10^{-3}$~particles~$h^3$~Mpc$^{-3}$, matching the densities of the real galaxy samples to be presented in \S~\ref{subsec:FilData}.  For each sample, we recompute the SHMAFF parameters and run the filament finder on all three smoothing scales, using the parameters derived in previous sections.  We will call a `false positive' any filament element found in the sparsely sampled data whose nearest neighboring element in the `true' filament network is more than a smoothing length away or does not have an axis of structure within an angle equal to $C \times l$.  Similarly, incompleteness is quantified by counting the filament elements in the `true' network that have no counterparts in the sparse-sampled one.

\begin{figure}[t]
\plotone{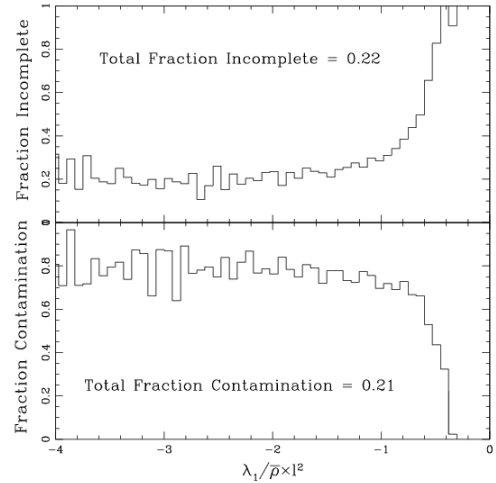}
\caption{Incompleteness and contamination of filament pixels as a function of $\lambda_1$ (scaled to the mean density and smoothing length) in the filament elements of a dark matter particle distribution sparse-sampled to match the density of $M_r<-20$ galaxies ($5.0 \times 10^{-3}$~h$^3$~Mpc$^{-3}$).  The sparse-sampled field was smoothed on a scale of $5$\hmpc, yielding $2.6$ particles per smoothing volume, and its filaments were compared with those in the full dark matter particle distribution smoothed on the same scale.  Their total incompleteness and contamination rates were both $\sim 20$~per~cent.\label{fig:SparseHistos}}
\end{figure}

As illustrated in Fig.~\ref{fig:SparseHistos}, the incompleteness and contamination rates of individual filament elements are strong functions of $\lambda_1$ for the `weakest' edges, but these make up only a small fraction of the filament network.  Our tests (not shown) suggest that the incompleteness and contamination rates are $\lesssim 20$~per~cent so long as there are an average of $\gtrsim 5$ particles within spheres of radius equal to the Gaussian smoothing length.  See \S~\ref{subsec:MockFils} for a more detailed analysis of completeness and contamination in mock galaxy samples.

\section{ Filaments as non-gaussianities   }
\label{subsec:FilGauss}

Gaussian random fields serve as an important reference point if we wish to distinguish the consequences of the non-linear growth of structure from phenomena seen only in the linear regime.  We know that Gaussian random fields are not filamentary and one might question why we should find {\it any} filaments in such a distribution.  Bear in mind, however, that the SHMAFF algorithm traces any negatively curved region and these conditions will certainly be met by some of the overdensities in a Gaussian random field.  We demonstrated in Paper~1 that although the smoothed $\lambda_1$ distributions appeared `filamentary' in both a Gaussian random field and the evolved dark matter distribution, the latter showed alignment between the axis of structure and these filamentary minima in $\lambda_1$.  The filament-finding algorithm enables us to follow the axis of structure and trace out individual large-scale structures in each distribution.

\begin{figure}[t]
\plotone{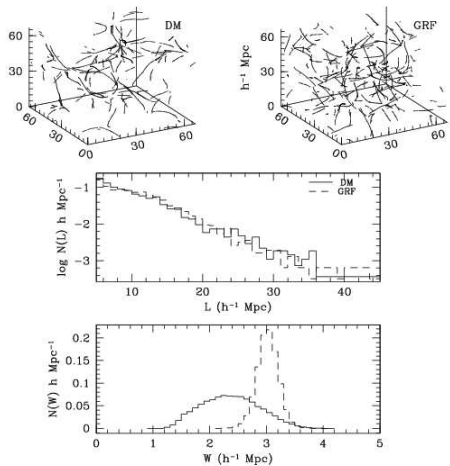}
\caption{Distribution of filaments in the dark matter distribution (no sparse sampling, upper left panel) and a Gaussian random field (upper right) with the $\Lambda$CDM $z=0$ non-linear power spectrum.  The filaments are found on $l=5$\hmpc~scales and we only plot subsections of the full $200$\hmpc~boxes.  In the centre and bottom panels, we show the filament length and width distributions, respectively, for the dark matter distribution (solid line) and Gaussian random field (dashed line).  In both cases, the filament-finding algorithm was run with $C=30$\ctunit~and $K=1$.\label{fig:FilsGRFCompare_all}}
\end{figure}

Using the $z=0$ linear power spectrum \citep{WMAP2} with corrections in the non-linear regime \citep[][the same one used in the N-body simulations discussed here]{NonLinearPS}, we generate a continuous realization of a three-dimensional Gaussian random field.  We use $l=5$\hmpc, $C=30$\ctunit, $K=1$ to derive the filament distribution shown in the upper right panel of Fig.~\ref{fig:FilsGRFCompare_all} and compare it with identical runs on the $z=0$ dark matter distribution in the cosmological simulation (upper left panel).  The qualitative differences between the two are substantial.  While the output for the dark matter distribution resembles a noded network, with filaments converging and ending at vertices in the network, the `filaments' in the Gaussian random field appear more randomly oriented and show no apparent correlations with one another.  Using $5$\hmpc~smoothing, the filament length distributions for the Gaussian random field and dark matter distribution are shown in the centre panel of Fig.~\ref{fig:FilsGRFCompare_all}.  The distributions are very similar and clearly exponential above a length of $\sim 10$\hmpc, with $N(L) \sim 10^{\frac{-0.1L}{\rm Mpc}}$, suggesting that filaments have not collapsed much along their longest axis since their formation, but have changed their alignment in relation to nearby structures.

We will define the width of a filament element, $W$, to be the root
mean squared perpendicular offset of particles within a
smoothing length; that is,
\begin{equation}
\label{eq:Width}
W=\sqrt{\frac{\sum_{i=1}^{N}|\mathbf{R}_i|^2}{N}},
\end{equation}
where $\mathbf{R_i}$ is defined in Equation~\ref{eq:PerpDisplace} and the sum is over all of the $N$ particles within one smoothing length of the filament element.  In the bottom panel of Fig.~\ref{fig:FilsGRFCompare_all}, we plot the width distributions for the two fields, again using $l=5$\hmpc.  The dark matter width distributions are broader and are peaked at smaller widths, suggesting that the filaments have collapsed significantly along two of their principal axes, despite having a similar length distribution.  As one would expect with bottom-up structure formation, the width distribution in the Gaussian random field and dark matter distribution are more discrepant at smaller smoothing scales (other scales not shown).

\subsection{ Filament evolution   }
\label{subsec:FilEvolve}

\begin{figure}[t]
\plotone{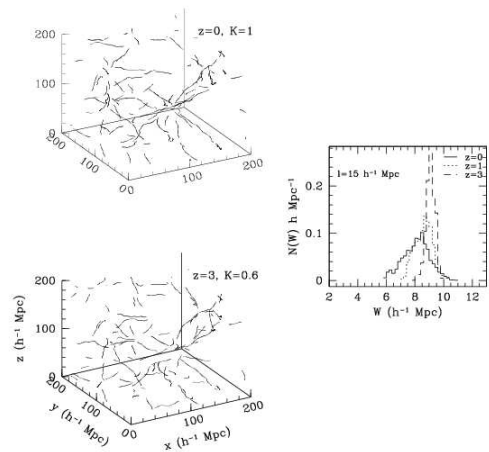}
\caption{Dark matter 
filament distributions at $z=3$ and $z=0$ after $l=15$\hmpc~smoothing.  The former was found with
a smaller value of $K$ because the algorithm tends to overestimate
filament widths when the filaments are of low contrast (see text).
Both the number and spatial distribution of filaments appear to be
similar at the two redshifts, but the width distributions are not, as
shown in the right panel.  Here, the filament width distributions at
$z=3$ (dashed line), $z=1$ (dotted line), and $z=0$ (solid line) are plotted.
\label{fig:EvolveAll}}
\end{figure}

In Paper~$1$, we showed that on a given comoving smoothing scale, there was evidence for a wall-to-filament-to-clump evolution with cosmic time.  Furthermore, we showed that the axis of structure aligns with the filamentary backbone in two-dimensional slices from cosmological simulations as early as $z=3$ (see figure~14 in Paper~1).  Fig.~\ref{fig:EvolveAll} shows the filament distribution at $z=0$ and $z=3$, now with $l=15$\hmpc\ so as to test the largest and least-evolved structures in the simulation box.  We used a smaller removal width, $K=0.6$, for the $z=3$ filament distribution because the filaments are of lower contrast than at $z=0$, causing Equation~\ref{eq:RemoveWidth} to overestimate their sizes.  The $z=3$ and $z=0$ filament distributions are very similar to the eye, suggesting that the basic filament framework for $l=15$\hmpc~is almost entirely in place at $z=3$ (where $15$\hmpc~fluctuations have $\left<\left(\Delta M/M\right)^2\right>^{1/2}\sim 0.1$).  The righthand panel of Fig.~\ref{fig:EvolveAll} shows the filament element width distributions as a function of redshift.  As non-linear evolution proceeds, the filament width distributions broaden and peak at smaller widths.

\section{ Filaments in the mock galaxy catalogues}
\label{subsec:MockFils}

Before we proceed to identify filaments in the SDSS data, we run the filament finder on the mock galaxy samples in redshift space (see Paper~1) and compare the resulting filaments to those identified in the real-space $z=0$ dark matter distribution.  The $l=5$\hmpc~filament distribution is very strongly affected by redshift distortions -- the contamination rates are typically $\sim 40$~per~cent, about double the contamination of the filament samples without redshift distortions.  This is due primarily to the `finger-of-god' effect, which causes galaxy clusters to extend into narrow, sharp filament-like features along the line of sight.  Fortunately, the filament finder is insensitive to these distortions on $10$\hmpc~and $15$\hmpc~scales because the fingers-of-god are typically $\lesssim 1$\hmpc~in width.  Nevertheless, we can improve our results if we first remove the fingers-of-god.

\subsection{ Identification and removal of fingers-of-god   }
\label{subsec:ClusterFind}

Fingers-of-god from galaxy clusters are extended along the
observer's line of sight, while real filamentary structure have no
preferred direction.  In order to separate the fingers-of-god from the 
real filaments, we will use a friends-of-friends algorithm
with two linking lengths,
\begin{eqnarray}
b_\parallel=\mathbf{b}\cdot\mathbf{\hat{r}} \nonumber \\
b_\perp=\sqrt{|\mathbf{b}|^2-b_\perp^2}
\label{eq:linlenZ}
\end{eqnarray}
where $\mathbf{\hat{r}}$ is the unit vector along the observer's line
of sight \citep{FOF}.  With these two parameters defined, the
algorithm searches for cylindrical structures with a
diameter-to-length ratio of $b_\perp/b_\parallel$.
\citet[e.g.][hereafter B06]{ZClusterFind2} did an exhaustive study of
this two-parameter space and found that $b_\perp=0.14$ and
$b_\parallel=0.75$ gave unbiased estimates of the group multiplicity
function, so we adopt these values in our study.

\begin{figure}[t]
\plotone{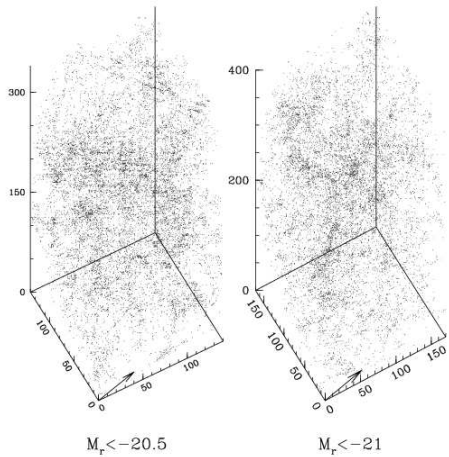}
\caption{Two volume--limited samples taken from the SDSS VAGC
large--scale structure sample, with galaxies placed at their comoving
positions based on the concordance cosmology.  The arrows indicate the location of the
Milky Way, which is $\vec{r}=(310,-20,170)$\hmpc~and $\vec{r}=(400,-25,200)$\hmpc~in the $Mr205$ and $Mr21$ samples, respectively.  The $z$ axes are parallel to the galactic north pole.\label{fig:DataBoxes}}
\end{figure}

\subsection{Filaments after cluster collapse}
\label{subsubsec:FilClustRemove}

All of the tests in this section were performed on the samples of mock galaxies with density similar to that of $M_r<-20$ galaxies.  First, we removed galaxy clusters from the {\it real space} mock galaxy distribution using an isotropic friends-of-friends algorithm with $b_{||}=b_{\perp}=0.2$ and a minimum group size of $N_{min}=5$.  For $l=10$\hmpc, the filament incompleteness and contamination rates for the cluster-free filament distribution ($33$~and $39$~per~cent, respectively) are much larger than those in real space  ($16$~and $25$~per~cent), suggesting that overdensities on megaparsec scales are playing an important role in defining filaments on $10$\hmpc\ scales.  Similar results are obtained when clusters are found and removed in redshift space using the approach of \S~\ref{subsec:ClusterFind}.

If we instead {\it collapse} the fingers-of-god presented by galaxy clusters, we can remove most of the contamination without having to remove the clusters themselves.  For this study, we will take the very simple approach of moving all members of a particular cluster to their mean position -- that is, we will collapse the fingers-of-god to a point weighted by number of galaxies in the cluster.  If we follow this procedure, the incompleteness and contamination are smaller ($19$~and $26$~per~cent) than the cluster-free mock galaxy distributions and a marginal improvement over the redshift space distribution with no special treatment of clusters  ($20$~and $27$~per~cent).

We repeated this exercise for filaments found on a $5$\hmpc~smoothing scale.  Collapsing the fingers-of-god does lead to a marginal improvement, but filaments are still very poorly defined in redshift space at these densities, with $\sim 40$~per~cent contamination rates.  A more sophisticated treatment of the clusters may be needed, but is beyond the scope of this paper.  In the section that follows we will discuss the application of the filament finder to real SDSS data.  To minimize contamination, we will be working only with filaments found on $10$\hmpc\ and $15$\hmpc\ scales and only after collapsing fingers-of-god.

\begin{figure}[t]
\plotone{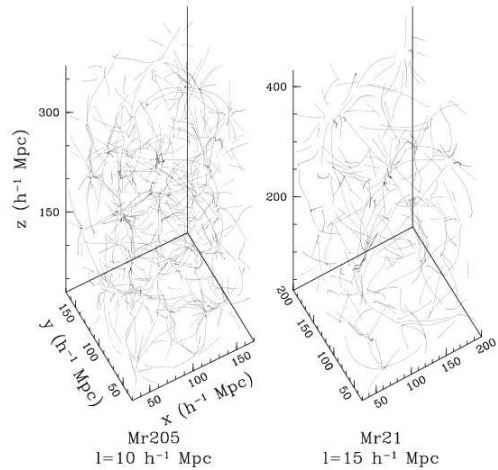}
\caption{Filaments found in the $Mr205$ (left, $l=10$\hmpc) and $Mr21$
(right, $l=15$\hmpc) SDSS samples.  These are the full filament
samples after the filament finder
was run with the `best' parameters.  Note that the boxes are
different (overlapping) volumes of space and thus are not directly
comparable to one another.
\label{fig:DataFils}}
\end{figure}

\section{ Filaments in the SDSS galaxy distribution }
\label{subsec:FilData}

The Sloan Digital Sky Survey has imaged a quarter of the sky in five wavebands, ranging from $3000$ to $10000$~\AA, to a depth of $r \sim 22.5$ \citep{SDSS}.  As of Data Release $6$ \citep{DR6}, spectra had been taken of $\sim 800,000$ galaxies, covering $9583$ square degrees and extending to Petrosian $r \sim 17.7$ \citep{Spectro}.  Galaxy redshifts are typically accurate to $\sim 30$~km~s$^{-1}$, making it ideal for studies of large--scale structure.  For this study, we need a portion of sky with relatively few coverage gaps to minimize the effect of the window function on the $\lambda$--space distributions.  With this in mind, we construct two volume--limited subsamples from the northern portion ($8<\alpha<16$~h and $25<\delta<60$) of the NYU Value-Added Galaxy Catalog \citep[NYU-VAGC, ][through DR$6$]{VAGC}, the first $140 \times 140 \times 340$~($h^{-1}$~Mpc)$^3$~in size with $M_r<-20.5$ ($Mr205$) and the second $170 \times 170 \times 400$~($h^{-1}$~Mpc)$^3$~in size with $M_r<-21$ ($Mr21$).  The samples extend to maximum redshifts of $z=0.12$ and $z=0.15$, respectively, and are plotted in redshift space in Fig.~\ref{fig:DataBoxes}.  Absolute magnitudes were computed with {\it kcorrect} \citep{Blanton03} using SDSS $r$-band Petrosian magnitudes shifted to $z=0.1$ (and using $h=1$).

We described the compilation and processing of the SDSS subsamples and their mock counterparts in Paper~$1$.  Before generating the filament distributions, we identify and collapse the fingers-of-god as described in the last section.  After performing this procedure on the $Mr205$ and $Mr21$ samples (both real and mock), we smooth the former with $l=10$\hmpc~and the latter with $l=15$\hmpc.  These choices maximize the volume covered while keeping the sampling rate high enough that filament contamination is under $\sim 25$~per~cent (see \S~\ref{subsubsec:FilClustRemove}).

We run the filament finder on the $Mr205$ and $Mr21$ galaxy samples using $C=40$\ctunit~and $C=50$\ctunit, respectively, and $K=1$.  The resulting filaments are shown in Fig.~\ref{fig:DataFils}.  After removing filaments shorter than a smoothing length, the algorithm finds $489$ filaments in $Mr205$, having a total length per unit volume of $1.9 \times 10^{-3}$~$h^{2}$~Mpc$^{-2}$ ($l=10$\hmpc), while in $Mr21$, $226$ filaments are found with a total length per unit volume of $7.6 \times 10^{-4}$~$h^{2}$~Mpc$^{-2}$ ($l=15$\hmpc).  For comparison, the mock $Mr205$ catalogue contains $451$ filaments with a total length per unit volume of $1.7 \times 10^{-3}$~$h^{2}$~Mpc$^2$ ($l=10$\hmpc) and the mock $Mr21$ catalogue contains $235$ filaments with a total length per unit volume of $8.2 \times 10^{-4}$~$h^{2}$~Mpc$^2$ ($l=15$\hmpc).  Thus, the number density of filaments in the simulations closely matches that in the real universe.

\begin{figure}[t]
\plotone{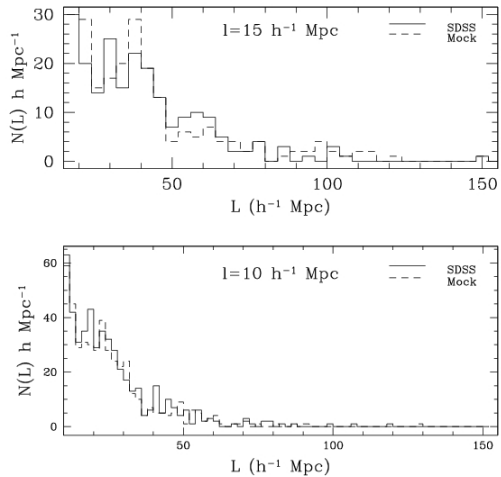}
\caption{Length distributions of the $Mr205$ (bottom) and $Mr21$ (top) filament samples, with SDSS filaments plotted with a solid line and redshift-space mock catalogues with a dashed line.\label{fig:DataLength}}
\end{figure}

We found in \S~\ref{subsec:FilGauss} that, above two smoothing lengths, dark matter filaments had an exponential length distribution that very closely matched that found in a Gaussian random field with the same power spectrum.  This suggests that, even if the filaments in the data are in a different stage of their evolution (i.e., having different $\sigma_8$) than those in the simulations, the length distributions should be the same between the two.  This does appear to be the case, as shown in Fig.~\ref{fig:DataLength}.

\begin{figure}[t]
\plotone{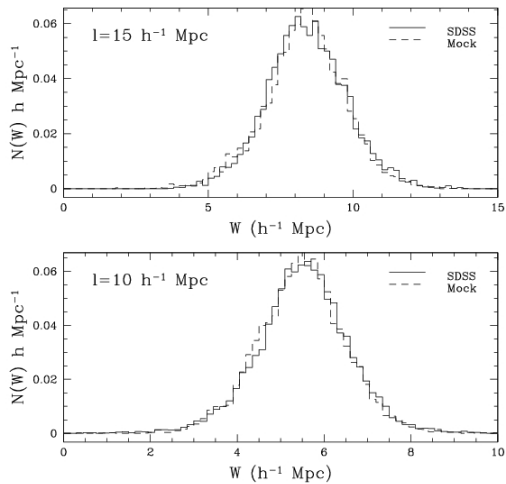}
\caption{Width distributions of the $Mr205$ (bottom) and $Mr21$ (top) filament samples, ith SDSS filaments plotted with a solid line and redshift-space mock catalogues with a dashed line.  The SDSS filaments in both samples are consistent with those in the cosmological simulations, suggesting that they are in similar stages of non-linear evolution.\label{fig:DataWidth}}
\end{figure}

More interesting is the similarity of the width distributions of filament elements, shown in Fig.~\ref{fig:DataWidth}.  In the SDSS, we find mean filament widths of $5.5$\hmpc\ and $8.4$\hmpc\ on $10$\hmpc\ and $15$\hmpc\ smoothing scales, with standard deviations of $1.1$\hmpc\ and $1.4$\hmpc, respectively.  As was demonstrated in Fig.~\ref{fig:EvolveAll}, filament element width distributions broaden and shift to smaller widths as non-linear evolution proceeds.  A large discrepancy in, for example, $\sigma_8$ between the simulations and real data should produce filament populations that are at different stages of non-linear evolution and have different width distributions.  As such, Fig.~\ref{fig:DataWidth} suggests that the SDSS filaments are both consistent with the standard model and consistent with the set of cosmological parameters used in the simulation.

\section{Results and discussion}
\label{subsec:Results}

This paper develops and uses an algorithm called the Smoothed Major
Axis Filament Finder to identify individual filaments in large-scale
structure.  In short, it uses the local eigenvectors of the
density second-derivative field to define the filament axis and trace individual
filaments.  Filament ends are defined as points at which the rate of
change of the axis of structure exceeds a specified threshold (see
\S~\ref{subsec:Method}).  In a $\Lambda$CDM cosmological simulation,
this definition produces filament samples that are consistent with our
visual impression of structure on a particular scale, are complete with
few duplicate detections (\S~\ref{subsubsec:Ct}), and are robust to sparse
sampling (\S~\ref{subsec:SparseFils}).  

In addition to the smoothing scale, the filament finder takes the input parameters $C$, the maximum angular rate of change of the filament axis, and $K$, the width of filament removal in units of the smoothing length.  Using Gaussian smoothing, the `best' values of these input parameters are $C=30$, $40$, and $50$\ctunit~on $5$, $10$, and $15$\hmpc~smoothing scales, respectively, and $K=1$ for all smoothing scales.  After we collapse the fingers-of-god, contamination and completeness in filament samples found in the mock $M_r<-20.5$ galaxy distribution are $\sim 26$ and $\sim 81$~per~cent, respectively.  Galaxy clusters are important for defining large-scale filaments and should not be removed before running the filament finder.  In redshift space and on smoothing scales above $\sim 10$\hmpc, collapsing fingers-of-god to their mean position produces mock filament samples comparable to those in real space.

In this paper, we presented two volume-limited subsamples from the northern portion of the SDSS spectroscopic survey (using the NYU-VAGC catalogue) and computed their filament distributions on $10$ and $15$\hmpc\ smoothing scales.  These distributions were then directly compared to those found in a series of redshift-space mock galaxy catalogues generated from a cosmological simulation using the concordance cosmology.  The filament length distributions found in SDSS data are very similar to those found in mock catalogues and are consistent with being drawn from an underlying exponential distribution.  The width distributions of filament elements are also very similar between the SDSS data and mock catalogues, suggesting that real filaments are consistent with those in a $\Lambda$CDM universe having $\sigma_8=0.85$, $\Omega_\Lambda=0.71$, $\Omega_m=0.29$, and $h=0.69$.  Tests on a range of cosmological simulations are needed before this can be turned into a cosmological constraint.

We also generated filament distributions at six redshifts in the
output of a $\Lambda$CDM cosmological N-body simulation, from $z=3$
to $z=0$.  The orientation of the filament network is stable out to
$z=3$ on comoving smoothing scales at least as large as $15$\hmpc.
Most of the filaments detected on $15$\hmpc~scales at $z=0$ can be
detected at $z=3$.  In addition, on a given comoving smoothing scale,
filament width distributions shift to smaller widths as the filaments continue to collapse.  Narrower filaments will collapse more rapidly, so this also leads to a broadening of the width distributions.

We have demonstrated that our filament finder is able to locate and
follow real structures, perhaps most strikingly in
\S~\ref{subsec:FilEvolve}, in which we showed that many of the same
structures could be seen in a cosmological simulation at both $z=3$
and $z=0$.  There is some subjective freedom in deciding what
constitutes the end of a filament, as no single physical threshold
stands out as a discriminator.  Nevertheless, we demonstrated in
\S~\ref{subsubsec:Ct} that the total length of the cosmic network is
insensitive to the choice of $C$ above a certain scale-dependent
threshold (once double detections are removed).  The minimum value of 
$C$ needed to probe the entire filament network may be telling us
about the {\it intrinsic} clumpiness of filamentary structure, and may 
therefore be able to distinguish models of warm and cold dark matter.

In this paper, we fully developed the SHMAFF algorithm and applied it to the low-redshift galaxy distribution, but there is much that can still be learned from its application to redshift surveys.  The filament evolution seen in cosmological simulations (see \S~\ref{subsec:FilEvolve}) can be tested in the DEEP2 galaxy survey \citep{DEEP2} at $z \sim 1$, and the results of this comparison have already been presented in \citet{EnaFil}.  On $l=5$\hmpc\ and $l=10$\hmpc\ scales, they confirm a shift in the filament width distribution to smaller widths from $z \sim 0.8$ to $z \sim 0.1$, as well as a broadening of the filament width distribution.  A possible extension of this work is a careful test of the $\Lambda CDM$ cosmological model, including precision constraints on cosmological parameters, such as $\sigma_8$, and tests for primordial non-Gaussianity using the length distribution of filamentary structures.   In addition, it would be useful to elaborate on the relationship of large-scale filaments to galaxy clusters and to explore the properties of galaxies in filaments relative to the general galaxy population.  Finally, it would be interesting to conduct a careful search for walls in SDSS.  Paper~$1$ hinted at their presence in the data, but they were only present at low contrast and the $\lambda$-space distributions were not optimal for identifying individual wall-like structures.

\section{Acknowledgments}
  
Funding for the SDSS and SDSS-II has been provided by the Alfred
P. Sloan Foundation, the Participating Institutions, the National
Science Foundation, the U.S. Department of Energy, the National
Aeronautics and Space Administration, the Japanese Monbukagakusho, the
Max Planck Society, and the Higher Education Funding Council for
England. The SDSS Web Site is http://www.sdss.org/.

The SDSS is managed by the Astrophysical Research Consortium for the
Participating Institutions. The Participating Institutions are the
American Museum of Natural History, Astrophysical Institute Potsdam,
University of Basel, University of Cambridge, Case Western Reserve
University, University of Chicago, Drexel University, Fermilab, the
Institute for Advanced Study, the Japan Participation Group, Johns
Hopkins University, the Joint Institute for Nuclear Astrophysics, the
Kavli Institute for Particle Astrophysics and Cosmology, the Korean
Scientist Group, the Chinese Academy of Sciences (LAMOST), Los Alamos
National Laboratory, the Max-Planck-Institute for Astronomy (MPIA),
the Max-Planck-Institute for Astrophysics (MPA), New Mexico State
University, Ohio State University, University of Pittsburgh,
University of Portsmouth, Princeton University, the United States
Naval Observatory, and the University of Washington.

\bibliographystyle{apj}
\bibliography{apj-jour,Bond062910}

\newcommand{\noopsort}[1]{}
\begin{thebibliography}{}

\bibitem[\protect\citeauthoryear{{Adelman-McCarthy} et~al.,}{{Adelman-McCarthy}
   et~al.}{2008}]{DR6}
{Adelman-McCarthy} J.~K.,  et~al., 2008, \apjs, 175, 297

\bibitem[\protect\citeauthoryear{{Arag{\'o}n-Calvo}, {Jones}, {van de Weygaert}
  \& {van der Hulst}}{{Arag{\'o}n-Calvo} et~al.}{2007}]{MMF}
{Arag{\'o}n-Calvo} M.~A.,  {Jones} B.~J.~T.,  {van de Weygaert} R.,    {van der
  Hulst} J.~M.,  2007, \aap, 474, 315

\bibitem[\protect\citeauthoryear{{Aragon-Calvo}, {Platen}, {van de Weygaert} \&
  {Szalay}}{{Aragon-Calvo} et~al.}{2008}]{SpineWeb}
{Aragon-Calvo} M.~A.,  {Platen} E.,  {van de Weygaert} R.,    {Szalay} A.~S.,
  2008, ArXiv:astro-ph/0809.5104

\bibitem[\protect\citeauthoryear{{Berlind} et~al.,}{{Berlind}
  et~al.}{2006}]{ZClusterFind2}
{Berlind} A.~A.,  et~al., 2006, \apjs, 167, 1

\bibitem[\protect\citeauthoryear{{Blanton} et~al.,}{{Blanton}
  et~al.}{2003}]{Blanton03}
{Blanton} M.~R.,  et~al., 2003, \aj, 125, 2348

\bibitem[\protect\citeauthoryear{{Blanton} et~al.,}{{Blanton}
  et~al.}{2005}]{VAGC}
{Blanton} M.~R.,  et~al., 2005, \aj, 129, 2562

\bibitem[\protect\citeauthoryear{{Bond}, {Kofman} \& {Pogosyan}}{{Bond}
  et~al.}{1996}]{CosmicWeb}
{Bond} J.~R.,  {Kofman} L.,    {Pogosyan} D.,  1996, \nat, 380, 603

\bibitem[\protect\citeauthoryear{{Bond}, {Strauss} \& {Cen}}{{Bond}
  et~al.}{2009}]{paper1}
{Bond} N.,  {Strauss} M.,    {Cen} R.,  2009, arXiv:astro-ph/0903.3601

\bibitem[\protect\citeauthoryear{{Cen} \& {Ostriker}}{{Cen} \&
  {Ostriker}}{1999}]{CO99}
{Cen} R.,  {Ostriker} J.~P.,  1999, \apj, 514, 1

\bibitem[\protect\citeauthoryear{{Choi}, {Bond} \& {Strauss}}{{Choi}
  et~al.}{2010}]{EnaFil}
{Choi} E.,  {Bond} N.~A.,    {Strauss} M.~A.,  2010, ArXiv Astrophysics
  e-prints

\bibitem[\protect\citeauthoryear{{Colberg}, {Krughoff} \& {Connolly}}{{Colberg}
  et~al.}{2005}]{Colberg05}
{Colberg} J.~M.,  {Krughoff} K.~S.,    {Connolly} A.~J.,  2005, \mnras, 359,
  272

\bibitem[\protect\citeauthoryear{{Colless} et~al.,}{{Colless}
  et~al.}{2001}]{2dF}
{Colless} M.,  et~al., 2001, \mnras, 328, 1039

\bibitem[\protect\citeauthoryear{{Colombi}, {Pogosyan} \&
  {Souradeep}}{{Colombi} et~al.}{2000}]{CPS00}
{Colombi} S.,  {Pogosyan} D.,    {Souradeep} T.,  2000, Physical Review
  Letters, 85, 5515

\bibitem[\protect\citeauthoryear{{Dave}, {Hellinger}, {Primack}, {Nolthenius}
  \& {Klypin}}{{Dave} et~al.}{1997}]{Dave97}
{Dave} R.,  {Hellinger} D.,  {Primack} J.,  {Nolthenius} R.,    {Klypin} A.,
  1997, \mnras, 284, 607

\bibitem[\protect\citeauthoryear{{Davis} et~al.,}{{Davis}
  et~al.}{2003}]{DEEP2}
{Davis} M.,  et~al., 2003, in {P.~Guhathakurta} ed., Society of Photo-Optical
  Instrumentation Engineers (SPIE) Conference Series Vol.~4834, {Science
  Objectives and Early Results of the DEEP2 Redshift Survey}.
pp 161--172

\bibitem[\protect\citeauthoryear{{Davis}, {Huchra}, {Latham} \&
  {Tonry}}{{Davis} et~al.}{1982}]{Davis82}
{Davis} M.,  {Huchra} J.,  {Latham} D.~W.,    {Tonry} J.,  1982, \apj, 253, 423

\bibitem[\protect\citeauthoryear{{de Lapparent}, {Geller} \& {Huchra}}{{de
  Lapparent} et~al.}{1986}]{CfaSurvey}
{de Lapparent} V.,  {Geller} M.~J.,    {Huchra} J.~P.,  1986, \apjl, 302, L1

\bibitem[\protect\citeauthoryear{{Dietrich}, {Schneider}, {Clowe},
  {Romano-D{\'{\i}}az} \& {Kerp}}{{Dietrich} et~al.}{2005}]{LensingFils2}
{Dietrich} J.~P.,  {Schneider} P.,  {Clowe} D.,  {Romano-D{\'{\i}}az} E.,
  {Kerp} J.,  2005, \aap, 440, 453

\bibitem[\protect\citeauthoryear{{Dolag}, {Meneghetti}, {Moscardini}, {Rasia}
  \& {Bonaldi}}{{Dolag} et~al.}{2006}]{Dolag06}
{Dolag} K.,  {Meneghetti} M.,  {Moscardini} L.,  {Rasia} E.,    {Bonaldi} A.,
  2006, \mnras, 370, 656

\bibitem[\protect\citeauthoryear{{Eisenstein} \& {Hut}}{{Eisenstein} \&
  {Hut}}{1998}]{HOP}
{Eisenstein} D.~J.,  {Hut} P.,  1998, \apj, 498, 137

\bibitem[\protect\citeauthoryear{{Forero-Romero}, {Hoffman}, {Gottl{\"o}ber},
  {Klypin} \& {Yepes}}{{Forero-Romero} et~al.}{2009}]{FR08}
{Forero-Romero} J.~E.,  {Hoffman} Y.,  {Gottl{\"o}ber} S.,  {Klypin} A.,
  {Yepes} G.,  2009, \mnras, 396, 1815

\bibitem[\protect\citeauthoryear{{Gonzalez} \& {Padilla}}{{Gonzalez} \&
  {Padilla}}{2009}]{GP09}
{Gonzalez} R.~E.,  {Padilla} N.~E.,  2009, arXiv:astro-ph/0912.0006

\bibitem[\protect\citeauthoryear{{Gott}, {Juri{\'c}}, {Schlegel}, {Hoyle},
  {Vogeley}, {Tegmark}, {Bahcall} \& {Brinkmann}}{{Gott}
  et~al.}{2005}]{SloanGreatWall}
{Gott} J.~R.~I.,  {Juri{\'c}} M.,  {Schlegel} D.,  {Hoyle} F.,  {Vogeley} M.,
  {Tegmark} M.,  {Bahcall} N.,    {Brinkmann} J.,  2005, \apj, 624, 463

\bibitem[\protect\citeauthoryear{{Hahn}, {Porciani}, {Carollo} \&
  {Dekel}}{{Hahn} et~al.}{2007a}]{Hahn07a}
{Hahn} O.,  {Porciani} C.,  {Carollo} C.~M.,    {Dekel} A.,  2007a, \mnras,
  375, 489

\bibitem[\protect\citeauthoryear{{Hahn}, {Carollo}, {Porciani} \&
  {Dekel}}{{Hahn} et~al.}{2007b}]{Hahn07b}
{Hahn} O.,  {Carollo} C.~M.,  {Porciani} C.,    {Dekel} A.,  2007b, \mnras,
  381, 41

\bibitem[\protect\citeauthoryear{{Hahn}, {Porciani}, {Dekel} \&
  {Carollo}}{{Hahn} et~al.}{2009}]{Hahn09}
{Hahn} O.,  {Porciani} C.,  {Dekel} A.,    {Carollo} C.~M.,  2009, \mnras, 398,
  1742

\bibitem[\protect\citeauthoryear{{Huchra} \& {Geller}}{{Huchra} \&
  {Geller}}{1982}]{FOF}
{Huchra} J.~P.,  {Geller} M.~J.,  1982, \apj, 257, 423

\bibitem[\protect\citeauthoryear{{Kaiser}, {Wilson}, {Luppino}, {Kofman},
  {Gioia}, {Metzger} \& {Dahle}}{{Kaiser} et~al.}{1998}]{Kaiser98}
{Kaiser} N.,  {Wilson} G.,  {Luppino} G.,  {Kofman} L.,  {Gioia} I.,  {Metzger}
  M.,    {Dahle} H.,  1998, ArXiv Astrophysics e-prints

\bibitem[\protect\citeauthoryear{{Massey} et~al.,}{{Massey}
  et~al.}{2007}]{LensingFils}
{Massey} R.,  et~al., 2007, \nat, 445, 286

\bibitem[\protect\citeauthoryear{{Moody}, {Turner} \& {Gott}}{{Moody}
  et~al.}{1983}]{GottFil}
{Moody} J.~E.,  {Turner} E.~L.,    {Gott} J.~R.~I.,  1983, \apj, 273, 16

\bibitem[\protect\citeauthoryear{{Novikov}, {Colombi} \& {Dor{\'e}}}{{Novikov}
  et~al.}{2006}]{Skeleton}
{Novikov} D.,  {Colombi} S.,    {Dor{\'e}} O.,  2006, \mnras, 366, 1201

\bibitem[\protect\citeauthoryear{{Pimbblet}}{{Pimbblet}}{2005}]{Pimbblet05a}
{Pimbblet} K.~A.,  2005, \mnras, 358, 256

\bibitem[\protect\citeauthoryear{{Platen}, {van de Weygaert} \&
  {Jones}}{{Platen} et~al.}{2007}]{Platen07}
{Platen} E.,  {van de Weygaert} R.,    {Jones} B.~J.~T.,  2007, \mnras, 380,
  551

\bibitem[\protect\citeauthoryear{{Press}, {Flannery} \& {Teukolsky}}{{Press}
  et~al.}{1986}]{NumericalRecipe}
{Press} W.~H.,  {Flannery} B.~P.,    {Teukolsky} S.~A.,  1986, {Numerical
  recipes. The art of scientific computing}

\bibitem[\protect\citeauthoryear{{Sahni}, {Sathyaprakash} \&
  {Shandarin}}{{Sahni} et~al.}{1998}]{Shapefinders}
{Sahni} V.,  {Sathyaprakash} B.~S.,    {Shandarin} S.~F.,  1998, \apjl, 495,
  L5+

\bibitem[\protect\citeauthoryear{{Sathyaprakash}, {Sahni}, {Shandarin} \&
  {Fisher}}{{Sathyaprakash} et~al.}{1998}]{Sathy98}
{Sathyaprakash} B.~S.,  {Sahni} V.,  {Shandarin} S.,    {Fisher} K.~B.,  1998,
  \apjl, 507, L109

\bibitem[\protect\citeauthoryear{{Sathyaprakash}, {Sahni} \&
  {Shandarin}}{{Sathyaprakash} et~al.}{1996}]{Sathy96}
{Sathyaprakash} B.~S.,  {Sahni} V.,    {Shandarin} S.~F.,  1996, \apjl, 462,
  L5+

\bibitem[\protect\citeauthoryear{{Schaap} \& {van de Weygaert}}{{Schaap} \&
  {van de Weygaert}}{2000}]{Schaap00}
{Schaap} W.~E.,  {van de Weygaert} R.,  2000, \aap, 363, L29

\bibitem[\protect\citeauthoryear{{Seldner}, {Siebers}, {Groth} \&
  {Peebles}}{{Seldner} et~al.}{1977}]{ShaneWirtanen}
{Seldner} M.,  {Siebers} B.,  {Groth} E.~J.,    {Peebles} P.~J.~E.,  1977, \aj,
  82, 249

\bibitem[\protect\citeauthoryear{{Sheth}, {Sahni}, {Shandarin} \&
  {Sathyaprakash}}{{Sheth} et~al.}{2003}]{SURFGEN}
{Sheth} J.~V.,  {Sahni} V.,  {Shandarin} S.~F.,    {Sathyaprakash} B.~S.,
  2003, \mnras, 343, 22

\bibitem[\protect\citeauthoryear{{Smith}, {Peacock}, {Jenkins}, {White},
  {Frenk}, {Pearce}, {Thomas}, {Efstathiou} \& {Couchman}}{{Smith}
  et~al.}{2003}]{NonLinearPS}
{Smith} R.~E.,  {Peacock} J.~A.,  {Jenkins} A.,  {White} S.~D.~M.,  {Frenk}
  C.~S.,  {Pearce} F.~R.,  {Thomas} P.~A.,  {Efstathiou} G.,    {Couchman}
  H.~M.~P.,  2003, \mnras, 341, 1311

\bibitem[\protect\citeauthoryear{{Sousbie}, {Pichon}, {Courtois}, {Colombi} \&
  {Novikov}}{{Sousbie} et~al.}{2008a}]{SkeletonData}
{Sousbie} T.,  {Pichon} C.,  {Courtois} H.,  {Colombi} S.,    {Novikov} D.,
  2008a, \apjl, 672, L1

\bibitem[\protect\citeauthoryear{{Sousbie}, {Pichon}, {Colombi}, {Novikov} \&
  {Pogosyan}}{{Sousbie} et~al.}{2008b}]{Skeleton3D}
{Sousbie} T.,  {Pichon} C.,  {Colombi} S.,  {Novikov} D.,    {Pogosyan} D.,
  2008b, \mnras, 383, 1655

\bibitem[\protect\citeauthoryear{{Spergel} et~al.,}{{Spergel}
  et~al.}{2007}]{WMAP2}
{Spergel} D.~N.,  et~al., 2007, \apjs, 170, 377

\bibitem[\protect\citeauthoryear{{Strauss} et~al.,}{{Strauss}
  et~al.}{2002}]{Spectro}
{Strauss} M.~A.,  et~al., 2002, \aj, 124, 1810

\bibitem[\protect\citeauthoryear{{Tanaka}, {Hoshi}, {Kodama} \&
  {Kashikawa}}{{Tanaka} et~al.}{2007}]{CL0016}
{Tanaka} M.,  {Hoshi} T.,  {Kodama} T.,    {Kashikawa} N.,  2007, \mnras, 379,
  1546

\bibitem[\protect\citeauthoryear{{Thompson} \& {Gregory}}{{Thompson} \&
  {Gregory}}{1978}]{TG78}
{Thompson} L.~A.,  {Gregory} S.~A.,  1978, \apj, 220, 809

\bibitem[\protect\citeauthoryear{{York} et~al.,}{{York}  et~al.}{2000}]{SDSS}
{York} D.~G.,  et~al., 2000, \aj, 120, 1579

\bibitem[\protect\citeauthoryear{{Zheng}, {Coil} \& {Zehavi}}{{Zheng}
  et~al.}{2007}]{HODPars2}
{Zheng} Z.,  {Coil} A.~L.,    {Zehavi} I.,  2007, \apj, 667, 760

\end{thebibliography}

\end{document}